\documentclass{iau}
\usepackage{graphicx}

\title[Rings of C$_2$H in Molecular Disks] 
{Rings of C$_2$H in the Molecular Disks \\ Orbiting TW Hya and V4046 Sgr}

\author[J. Kastner et  al.]   
{J. H. Kastner$^1$, C. Qi$^2$, U. Gorti$^3$, P. Hily-Blant$^4$, K. Oberg$^2$, T. Forveille$^4$, S. Andrews$^2$, and D. Wilner$^2$}

\affiliation{$^1$Lab.\ for Multiwavelength Astrophysics, Rochester Institute of Technology; {\tt jhk@cis.rit.edu}\\
  $^2$ Harvard-Smithsonian Center for Astrophysics\\
  $^3$ SETI Institute \& NASA Ames Research Center\\
  $^4$ Universit\'e Grenoble Alpes \& Institut de Plan\'etologie et dÕAstrophysique de Grenoble
}

\pubyear{2015}
\volume{314}  
\pagerange{119--126}
\jname{Young Stars \& Planets Near the Sun}
\editors{J. H. Kastner, B. Stelzer, \& S. A. Metchev, eds.}
\begin{document}

\maketitle

\begin{abstract}
  We have used the Submillimeter Array to image, at $\sim$1$''$
  resolution, C$_2$H(3--2) emission from the molecule-rich
  circumstellar disks orbiting the nearby, classical T Tauri star
  systems TW Hya and V4046 Sgr. The SMA imaging reveals that the
  C$_2$H emission exhibits a ring-like morphology within each disk; the
  the inner hole radius of the C$_2$H ring within the V4046 Sgr disk
  ($\sim$70 AU) is somewhat larger than than of its counterpart within
  the TW Hya disk ($\sim$45 AU).  We suggest that, in each case, the
  C$_2$H emission likely traces irradiation of the tenuous surface
  layers of the outer disks by high-energy photons from the central
  stars.  
\keywords{protoplanetary disks, stars: individual
    (TW Hya, V4046 Sgr), stars: pre-main sequence}
\end{abstract}

\firstsection 
\section{Introduction}

Our single-dish radio telescope molecular line surveys of the evolved
protoplanetary disks orbiting the nearby, actively accreting T Tauri
star systems TW Hya and V4046~Sgr established that both display strong
C$_2$H line emission (\cite[Kastner et  al.\ 2014]{2014ApJ...793...55K}). As C$_2$H is likely the
dissociation product of more complex hydrocarbons and organics, it
potentially provides a diagnostic of disk irradiation by high-energy
(UV and X-ray) photons generated via accretion and coronal activity at
the central T Tauri stars. With this as motivation, we have used the
Submillimeter Array (SMA) to image, at $\sim$1$''$ resolution,
C$_2$H(3--2) emission from the TW Hya and V4046 Sgr disks.

\vspace{-.15in}
\section{SMA observations of C$_2$H emission from TW Hya and V4046
  Sgr}

\begin{figure}[h!]
\begin{center}
\includegraphics[width=5in]{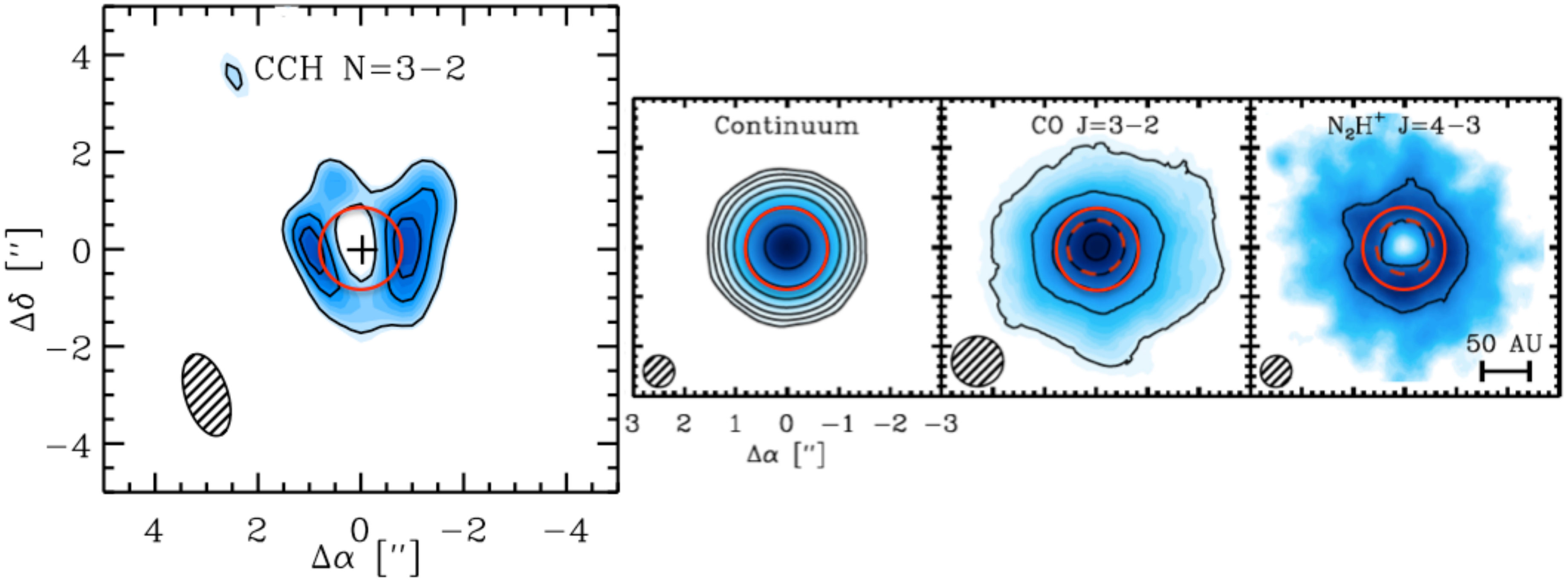}
\caption{Interferometric imaging of the TW Hya disk. Left panel: SMA
  C$_2$H(3--2) image, from K15. Left center, right center, and right
  panels: ALMA 372 GHz continuum, SMA CO(3--2), and ALMA
  N$_2$H$^+$(4--3) images, from \cite[Qi et al.\ (2013)]{2013Sci...341..630Q}. The red circles
  indicate the $\sim$45 AU inner radius of the C$_2$H ring, as determined from model
  fitting described in K15.  }
\label{fig:TWHyaC2H}
\end{center}
\end{figure}

{\bf TW Hya:} The results of SMA imaging of C$_2$H emission from this
single, nearby ($D=54$ pc) $\sim$0.8 $M_\odot$ T Tauri star --- the
namesake of the $\sim$8 Myr-old TW Hya Association --- are reported in
\cite[Kastner et al.\ (2015; hereafter K15)]{2015arXiv150405980K}. These observations have revealed
that the C$_2$H line surface brightness exhibits a ring-like emission
morphology, with an inner hole of radius $\sim$45 AU (Fig.~\ref{fig:TWHyaC2H}).  Based on comparison of the SMA
C$_2$H imaging results for TW Hya with previous continuum and line
emission imaging and spectroscopy of that disk, it appears that the
C$_2$H abundance is enhanced in warm, low-density, disk surface layers
that lie beyond the CO ``ice line'' (as traced by N$_2$H$^+$ emission;
Fig.~\ref{fig:TWHyaC2H}).

\begin{figure}[h!]
\begin{center}
\includegraphics[width=3.5in]{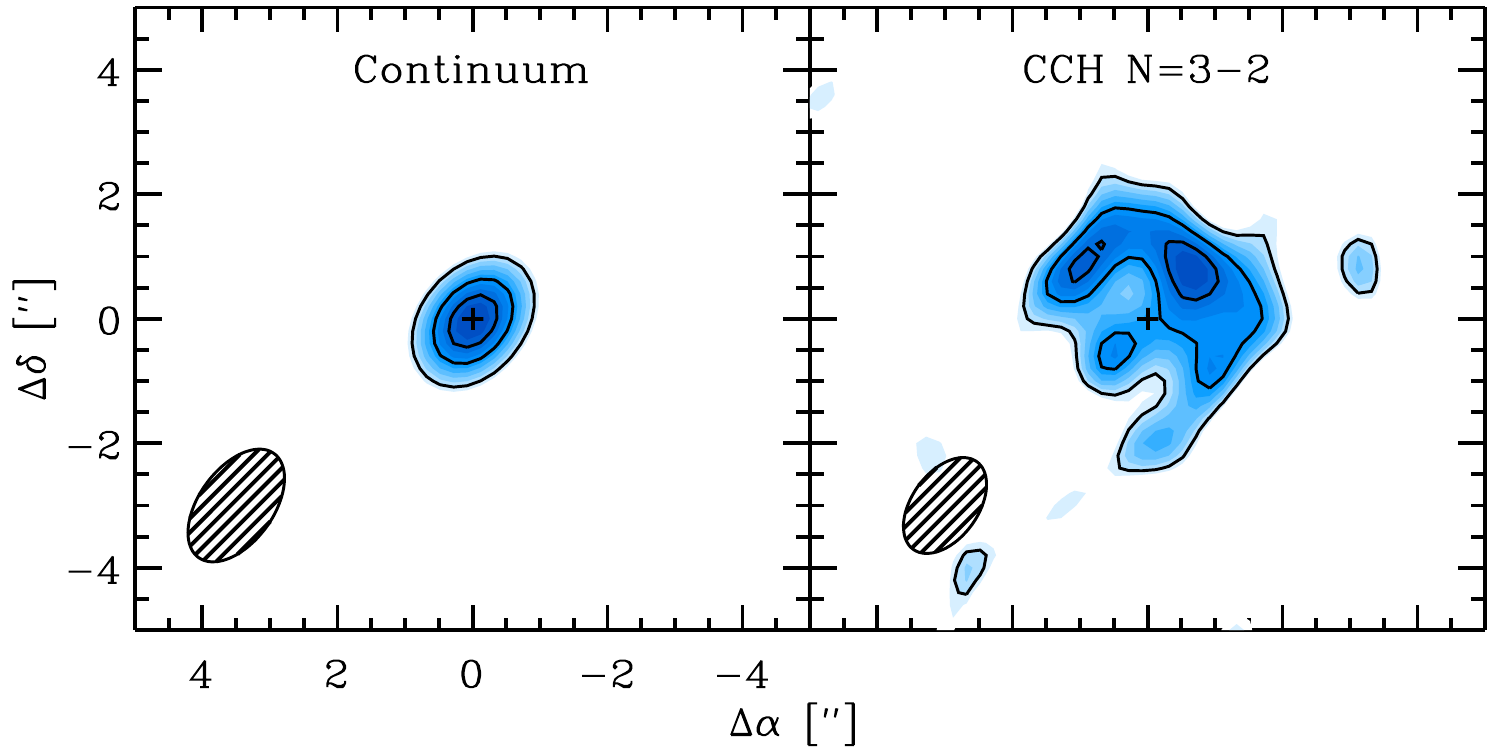}
\caption{SMA C$_2$H(3--2) image of the disk orbiting V4046 Sgr (right)
  alongside the 260 GHz continuum image obtained from the same
  (extended+compact configuration) dataset (left). }
\label{fig:V4046SgrC2H}
\end{center}
\vspace{-.25in}
\end{figure}

{\bf V4046 Sgr:} This close ($\sim$2.4-day) binary system, a member of
the $\sim$20 Myr-old $\beta$ Pic Moving Group, consists of nearly
equal, $\sim$0.9 $M_\odot$ components (\cite[Rosenfeld et al.\ 2012]{2012ApJ...759..119R}).  The circumbinary molecular disk orbiting V4046 Sgr
is somewhat larger than that of TW Hya ($R \sim 350$ AU vs.\ $R \sim
200$ AU; \cite[Rosenfeld et al.\ 2012]{2012ApJ...759..119R}; \cite[Qi
et al.\ 2013]{2013Sci...341..630Q}), and --- in
contrast to the nearly pole-on TW Hya --- is viewed at intermediate
inclination ($i \approx 34^\circ$; Rosenfeld et al.\ 2012). Extended-
and compact-configuration SMA observations of C$_2$H(3--2) emission
from the V4046 Sgr disk were acquired in 2015 March and May,
respectively; the resulting image is displayed in Fig.~\ref{fig:V4046SgrC2H}. It is evident that V4046
Sgr, like TW Hya, displays a ring-like C$_2$H morphology, but with a
more asymmetric appearance (perhaps due to its intermediate disk
inclination) and
somewhat larger inner C$_2$H ``hole'' radius ($\sim$70 AU).

\section{Remarks}

We (K15) have proposed that the C$_2$H ring in the TW Hya disk may
trace particularly efficient photo-destruction of small grains and/or
photodesorption and photodissociation of hydrocarbons (principally,
C$_2$H$_2$) derived from grain ice mantles in the surface layers of
the outer disk. Comparative modeling of the two SMA datasets will
establish whether the V4046 Sgr C$_2$H ring is in
fact larger and more asymmetric than that of TW Hya and, more generally, should shed further
light on the mechanism(s) responsible for production of C$_2$H in evolved, protoplanetary disks.

\end{document}